\documentclass[12pt]{article}
\usepackage{epsfig}
\begin{document}
\noindent \thispagestyle{empty}
\begin{center}
{\bf \large Taking into Account the Centre-of-Mass Correlations\\
in the Cross Sections for Elastic Scattering
of Intermediate Energy Protons
on the Exotic Nuclei $^6$He and $^8$He} \\
\end{center}
\vspace{3mm}
\begin{center}
{\bf G.~D.~Alkhazov and V.~V.~Sarantsev}
\end{center}
\begin{center}
{\it National Research Centre ``Kurchatov Institute'' --\\
Petersburg Nuclear Physics Institute, Gatchina, Russia}\\
\end{center}
\setcounter{page}{1}
{\small {\bf Abstract}---We calculate the differential cross sections for proton elastic scattering on the exotic halo nuclei $^6$He
and $^8$He at energies around $\sim$0.7 GeV at the
momentum transfers squared up to 0.30 (GeV/$c$)$^2$
and investigate the influence of the nucleon centre-of-mass
correlations on the calculated cross sections. In particular,
we show that the approximate account of the centre-of-mass
correlations used previously considerably overestimates
the cross sections at high values of the momentum transfer.}\\
{\bf Comments:} 12 pages, 2 figures. Submitted to Physics of Atomic Nuclei\\
{\bf Subjects:} Nuclear Theory(nucl-th); High-Energy Physics-Theory(hep-th)\\
\begin{center}
1. INTRODUCTION\\
\end{center}

Proton-nucleus elastic scattering at intermediate energy is an
efficient means of studying the nuclear spatial structure.
Several experiments were performed [1--3] in which the cross
sections for proton elastic scattering on light exotic nuclei
were measured in inverse kinematics at GSI-Darmstadt at an energy
of $\sim$700 MeV/u at small momentum transfers ($\sim$0.002 $\leq|t|\leq \hspace*{3pt}\sim$0.05 (GeV/$c$)$^2$,
where $t$ is the four momentum transfer squared). The cross sections
were measured with the help of the hydrogen-filled ionization
chamber IKAR [4], which served simultaneously as a gas target
and a detector of the recoil protons. In particular, the
$p^6$He and $p^8$He cross sections were measured [1].

The measured cross sections were analysed [5] in the framework of
the Glauber multiple-scattering theory, and the
root-mean-square (rms) radii of the
nuclear total matter $R_{\rm m}$, the nuclear core $R_{\rm c}$
and the neutron halo $R_{\rm h}$ were deduced for $^6$He and $^8$He. Later, the $p^6$He and $p^8$He cross sections for elastic scattering
were \mbox{measured [6]} practically at the same energy at higher momentum
transfers (up to $|t|$ = 0.225 (GeV/$c$)$^2$) using an experimental set-up with a liquid hydrogen target. The new
experimental data at higher values of $|t|$ in combination with the cross sections [1] measured at low $|t|$-values allow, in principle, to determine the $^6$He and $^8$He nuclear radii $R_{\rm c}$, $R_{\rm h}$ and $R_{\rm m}$
with better accuracy [7].
The many-body density distributions of $^6$He and $^8$He used
in the calculations [7] of the cross sections were represented as
products of one-body densities. The effect of the centre-of-mass
(CM) correlations was taken into account by an approximate
approach (to be considered below).

As is known, the effect of the CM correlations in the calculated
cross sections is rather sizeable in the case of proton
scattering on light nuclei at large momentum transfers. In the
present paper, in order to find out the effect
of the CM correlations on the $p^6$He and $p^8$He cross sections,
we have calculated these cross sections neglecting the
CM correlations and taking them into account by approximate
and exact methods. \\

\begin{center}
 2. BASIC FORMULAS\\
\end{center}

According to the Glauber theory [8], the proton-nucleus elastic
scattering amplitude $F({\bf q})$ can be calculated as
\begin{displaymath}
F({\bf q}) = \frac{ik}{2\pi} \int d^2b ~ {\rm
exp} ~(i{\bf qb}) ~ \rho _A({\bf r}_1,...,{\bf r}_A) ~
d^3r_1 d^3r_2...d^3r_A ~ \times
\end{displaymath}
\begin{equation}
\times  \left\{1 - \prod_{j=1}^A \left[1 - \gamma({\bf b} - {\bf
s}_j) \right] \right\} .
\end{equation}
Here, $k$ is the value
of the wave vector of the incident proton,
{\bf q} is the momentum transfer, ${\bf b}$ is the impact-parameter
vector, $\rho _A({\bf r}_1,...,{\bf r}_A)$ is the many-body density
distribution of the nucleus, ${\bf r}_1,...,{\bf r}_A$ and
${\bf s}_1,...,{\bf s}_A$ stand for the radius vectors of the
nucleons in the nucleus and their transverse coordinates,
$A$ is the total number of nucleons in the nucleus, and
$\gamma({\bf b})$ is the profile function of the nucleon--nucleon
interaction. We employ a spin-independent amplitude of the
free proton--nucleon ($pN$) scattering with the
traditional high-energy parametrization of this amplitude and the
corresponding profile-function
\begin{equation}
\gamma({\bf b}) = \frac{\sigma_{pN}(1 - i\epsilon_{pN})}{2}  ~
\frac{1}{2 \pi \beta_{pN}} ~ {\rm exp}\Big(- \frac{{\bf b}^2}{2
\beta_{pN}}\Big),
\end{equation}
where $\sigma_{pN}$ is the total cross section for the $pN$ interaction,
$\beta_{pN}$ is the $pN$ amplitude slope parameter, and
$\epsilon_{pN}$ is the ratio of the real to imaginary parts
of the $pN$ scattering amplitude. The differential cross section
$d\sigma$/$dt$ for proton-nucleus elastic scattering is
connected with the amplitude $F({\bf q})$ as
\begin{equation}
d{\sigma}/ dt = (\pi/k^2)|F({\bf q})|^2.
\end{equation}

 If we neglect all nucleon correlations in the nuclear many-body
 density distribution $\rho _A({\bf r}_1,...,{\bf r}_{A})$, then it can be
 represented as a product of similar
 one-body densities $\rho_0 ({\bf r}_j)$:
\begin{equation}
\rho _A({\bf r}_1,...,{\bf r}_{A}) =
\prod_{j=1}^A \rho_0 ({\bf r}_j).
\end{equation}
Let $F_0({\bf q})$ be the proton-nucleus elastic scattering
amplitude calculated by \mbox{Eq. (1)} with many-body density (4).
Then in the case where the one-body density
$\rho_0 ({\bf r}_j)$ is a Gaussian distribution
\begin{equation}
\rho_{0} ({\bf r}_j) = (2\pi a_{0}^2)^{- 3/2} ~ {\rm
exp} (- {{\bf r}_j}^2 /2 a_{0}^2),
\end{equation}
the CM correlations can be taken into account
in the calculated amplitude $F_1({\bf q})$ as it is known [9]
by multiplying the amplitude $F_0({\bf q})$ with a
CM correction factor $H_0({\bf q})$:
\begin{equation}
F_1({\bf q}) = H_0({\bf q})F_0({\bf q}),
\end{equation}
\begin{equation}
\textrm{where} \hspace{3mm}
H_0({\bf q}) = ~ {\rm exp}[{\bf q}^2 R_{\rm m}^2/6 (A - 1)]
\end{equation}
and $R_{\rm m} \equiv <{\bf r^2}>^{1/2} = [3 (1 - 1/A)]^{1/2} a_0$
is the nuclear rms matter radius calculated in the nucleus
CM system.

Equation (6) with correction factor (7) takes into
account the CM correlations exactly in the case of a Gaussian
one-body nuclear matter density distribution \mbox{(Eq. (5)).}
For non-Gaussian distributions such an approach is approximate
and it is justified for the cases when the distributions of the total
nuclear density do not differ significantly from a Gaussian one and for
proton scattering on middle-weight and
heavy nuclei at relatively small momentum transfers where the
effect of the CM correlations is small.
However, such a correction factor was used in \mbox{[2, 3, 5, 7]}
for proton elastic scattering on light exotic halo nuclei,
the total matter density distributions in which differ significantly
from a Gaussian one.

In the following, in order to see how big is
the effect of the CM correlations in the cross sections and how
accurate is the approximate account of the correlations,
we perform calculations of
the cross sections for proton elastic scattering on the exotic
nuclei $^6$He and $^8$He at an energy of $\sim$700 MeV neglecting
the CM correlations, taking them into account exactly, and using
an approximate approach with the correction factor given by Eq. (7).

If we neglect the CM correlations in the nuclear many-body density
$\rho _A({\bf r}_1,...,{\bf r}_A)$, then for exotic nuclei
$^6$He and $^8$He it can be represented as
\begin{equation}
\rho _A({\bf r}_1,...,{\bf r}_A) = \prod_{j=1}^{A_{\rm c}} \rho_{\rm
c} ({\bf r}_j) \times \prod_{j=A_{\rm c} + 1}^{A_{\rm c} + A_{\rm
h}} \rho_{\rm h} ({\bf r}_j).
\end{equation}
Here $\rho_{\rm c} ({\bf r}_j)$ and $\rho_{\rm h} ({\bf r}_j)$
are the one-body density distributions of the core and halo of the exotic
nucleus correspondingly; $A_{\rm c}$ and $A_{\rm
h}$ are the numbers of the core and halo nucleons ($A_{\rm c} +
A_{\rm h} = A$; \mbox{$A_{\rm c} = 4$} in $^6$He and $^8$He;
$A_{\rm h} = 2$ in $^6$He and $A_{\rm h} = 4$ in $^8$He;
the core consists of 2 protons and 2 neutrons,
the halo consists of neutrons).
We describe the matter density distributions
in the core and in the halo by Gaussian functions
\begin{equation}
\rho_{\rm c} ({\bf r}_j) = (2\pi a_{\rm c}^2)^{- 3/2} ~
{\rm exp} (- {{\bf r}_j}^2 /2 a_{\rm c}^2),
\end{equation}
\begin{equation}
\rho_{\rm h} ({\bf r}_j) = (2\pi a^2_{\rm h})^{- 3/2} ~
{\rm exp} (- {{\bf r}_j}^2 /2 a_{\rm h}^2),
\end{equation}
where $a_{\rm c} = R_{\rm c}/\sqrt{3}$, $a_{\rm h} =
R_{\rm h}/\sqrt{3}$, $R_{\rm c}$ and $R_{\rm h}$
are the rms radii of the core and halo
matter density distributions (we assume $R_{\rm h} > R_{\rm c}$). As for the rms radius $R_{\rm m}$ of the total matter density
distribution, it is connected with the core and halo
radii $R_{\rm c}$ and $R_{\rm h}$ as
\begin{equation}
R_{\rm m} = \left[(A_{\rm c}R_{\rm c}^2 + A_{\rm h}R_{\rm h}^2)/A
\right]^{1/2}.
\end{equation}

Amplitude (1) calculated with many-body density
distribution (8) does not include the CM correlations.
If this amplitude is multiplied by correction factor (7),
then the CM correlations are taken into account
approximately. Note that in the latter case the parameters
$a_{\rm c}$ and $a_{\rm h}$ are connected with the core and
halo radii $R_{\rm c}$ and $R_{\rm h}$ as
\begin{equation}
a_{\rm c}^2 = [R_{\rm c}^2 + R_{\rm m}^2/(A - 1)]/3,
\end{equation}
\begin{equation}
a_{\rm h}^2 = [R_{\rm h}^2 + R_{\rm m}^2/(A - 1)]/3.
\end{equation}

In the case when the core and halo distributions are
Gaussian ones, the CM correlations can be taken
into account exactly in a simple way as described below.
We assume that the core of $^6$He and $^8$He consists
of a four-nucleon cluster with the size $R^*_{\rm c}$
the same as that of the $^4$He nucleus
($R^*_{\rm c}$ $\approx$ 1.46 fm [10]). This cluster
experiences some motion around the CM of the nucleus
so that the effective core size $R_{\rm c}$ is larger
than $R^*_{\rm c}$. Similarly, the effective halo size
$R_{\rm h}$ is larger than the size $R^*_{\rm h}$ of
the halo cluster (2 neutrons in $^6$He and 4 neutrons
in $^8$He) in its CM system. It is easy to show
(see, for example, [7]) that
\begin{equation}
R^*_{\rm h} = [R^2_{\rm h} - (A_{\rm c}/A_{\rm h})^2
(R^2_{\rm c} - {R^*_{\rm c}}^2)]^{1/2}.
\end{equation}

We start the calculation of the proton-nucleus
scattering amplitude $F_{A}({\bf q})$ by calculations
of the amplitudes $F^*_{\rm c}({\bf q})$ and
$F^*_{\rm h}({\bf q})$ of proton scattering on the core
and halo clusters using the Glauber formula (1). The
many-body density distributions of the core and halo
clusters are represented as products of one-body
density distributions $\rho ^*_{\rm c}({\bf r}_j)
\sim {\rm exp} (- {\bf r}^2_j / 2 {a^*_{\rm c}}^2)$ and
$\rho ^*_{\rm h}({\bf r}_j) \sim {\rm exp}
(- {\bf r}^2_j / 2 {a^*_{\rm h}}^2)$ where the parameters
$a^*_{\rm c}$ and $a^*_{\rm h}$ are connected with
the rms matter radii $R^*_{\rm c}$ and $R^*_{\rm h}$ as
\begin{equation}
a^*_{\rm c} = R^*_{\rm c}/[3(1 - 1/A_{\rm c})]^{1/2},
\end{equation}
\begin{equation}
a^*_{\rm h} = R^*_{\rm h}/[3(1 - 1/A_{\rm h})]^{1/2}.
\end{equation}
In order to take into account the CM nucleon correlations
in the core and halo clusters, the calculated amplitudes
are multiplied by the correction factors $H^*_{\rm c}({\bf q})$
and $H^*_{\rm h}({\bf q})$, where
\begin{equation}
H^*_{\rm c}({\bf q}) = {\rm exp}[{\bf q}^2 {R^*_{\rm c}}^2/6 (A_c - 1)],
\end{equation}
\begin{equation}
H^*_{\rm h}({\bf q}) = {\rm exp}[{\bf q}^2 {R^*_{\rm h}}^2/6 (A_h - 1)].
\end{equation}

Then we prescribe some artificial independent motion
of these clusters in the laboratory system in accordance
with the distributions $\rho_{\rm c}({\bf r}_{\rm c})$
and $\rho_{\rm h}({\bf r}_{\rm h})$:
\begin{equation}
\rho_{\rm c}({\bf r}_{\rm c}) = (2 \pi a^2_{\rm c,h}/A_{\rm c})^{- 3/2}
{\rm exp} (- A_{\rm c} {\bf r}^2_{\rm c} / 2 a^2_{\rm c,h}),
\end{equation}
\begin{equation}
\rho_{\rm h}({\bf r}_{\rm h}) = (2 \pi a^2_{\rm c,h}/A_{\rm h})^{- 3/2}
{\rm exp} (- A_{\rm h} {\bf r}^2_{\rm h} / 2 a^2_{\rm c,h}),
\end{equation}
\begin{displaymath}
\textrm{where}~
{\bf r}_{\rm c} = \sum_{j=1}^{A_{\rm c}} {\bf r}_j /A_{\rm c},
~
{\bf r}_{\rm h} = \hspace*{-2mm} \sum_{j=A_{\rm c} + 1}^{A_{\rm c} +
A_{\rm h}} {\bf r}_j /A_{\rm h},~
\textrm{and $a_{\rm c,h}$ is a radial parameter
to be defined}
\end{displaymath}
later. Correspondingly, the two-body density distribution
$\rho_{\rm c,h}({\bf r}_{\rm c},
{\bf r}_{\rm h})$ describing the core and halo clusters motion
in the laboratory system is
\begin{equation}
\hspace{-1 pt}\rho_{\rm c,h}({\bf r}_{\rm c},{\bf r}_{\rm h}) =
\rho_{\rm c}({\bf r}_{\rm c}) \cdot \rho_{\rm h}({\bf r}_{\rm h})
\sim {\rm exp}[- (A_c {\bf r}^2_c + A_h {\bf r}^2_h)/2 a^2_{\rm c,h}].
\hspace{-3 mm}
\end{equation}

We can calculate the amplitude $F_{\rm c,h}({\bf q})$
of proton scattering on the system of these clusters by
the basic Glauber formula using the amplitudes
$F^*_{\rm c}({\bf q})$ and $F^*_{\rm h}({\bf q})$, the two-body density distribution
$\rho_{\rm c,h}({\bf r}_{\rm c}, {\bf r}_{\rm h})$ and performing
integrations over the radius-vectors ${\bf r}_{\rm c}$ and
${\bf r}_{\rm h}$.
We obtain
\begin{equation}
F_{\rm c,h}({\bf q}) = \frac{ik}{2\pi}\int d^2b ~
{\rm exp} (i {\bf qb}) ~
\{1 - [1 - \Gamma _{\rm c} ({\bf b})] [1 - \Gamma _{\rm h}({\bf b})] \},
\end{equation}
where
\begin{equation}
\Gamma _{\rm c} ({\bf b}) = \frac{1}{2 \pi ik}
\int d^2 q ~ {\rm exp} (- q^2 a^2_{\rm c,h} / 2 A_{\rm c}) ~ F^*_{\rm c}({\bf q}) ~ {\rm exp} (-i {\bf qb}),
\end{equation}
\begin{equation}
\Gamma _{\rm h} ({\bf b}) = \frac{1}{2 \pi ik}
\int d^2 q ~ {\rm exp} (- q^2 a^2_{\rm c,h} / 2 A_{\rm h}) ~ F^*_{\rm h}({\bf q}) ~ {\rm exp} (-i {\bf qb}).
\end{equation}

Now we note that due to the following equality
\begin{equation}
A_{\rm c} {\bf r}^2_{\rm c} + A_{\rm h} {\bf r}^2_{\rm h} =
(A_{\rm c} {\bf r}'^2_{\rm c} + A_{\rm h} {\bf r}'^2_{\rm h})
+ A {\bf r}_{\rm CM}^2
\end{equation}
the two-body density
distribution $\rho_{\rm c,h}({\bf r}_{\rm c}, {\bf r}_{\rm h})$
(given by Eq. (21)) can be represented as a product of the distribution
$\rho ({\bf r}_{\rm CM})$ of the CM radius
${\bf r}_{\rm CM} = (A_{\rm c}
{\bf r_{\rm c}} + A_{\rm h} {\bf r_{\rm h}}) / A$ and the
distribution $\rho ' ({\bf r}_{\rm c}', {\bf r}_{\rm h}')$
of the relative coordinates ${\bf r}_{\rm c}' = {\bf r}_{\rm c}
- {\bf r}_{\rm CM}$ and ${\bf r}_{\rm h}' = {\bf r}_{\rm h}
- {\bf r}_{\rm CM}$:
\begin{equation}
\rho_{\rm c,h}({\bf r}_{\rm c}, {\bf r}_{\rm h}) =
\rho ({\bf r}_{\rm CM}) \cdot \rho ' ({\bf r}_{\rm c}', {\bf r}_{\rm h}'),
\end{equation}
\begin{equation}
{\textrm{where}} \hspace{3mm} \rho ({\bf r}_{\rm CM}) \sim {\rm exp}
(- A {\bf r}^2_{\rm CM} / 2 a^2_{\rm c,h})
\end{equation}
\begin{equation}
{\textrm{and}} \hspace{3mm} \rho ' ({\bf r}_{\rm c}', {\bf r}_{\rm h}')
\sim {\rm exp} [- (A_{\rm c} {\bf r}'^2_{\rm c} +
A_{\rm h} {\bf r}'^2_{\rm h}) / 2 a^2_{\rm c,h}].
\end{equation}

Let us calculate the amplitude $F_{\rm c,h}({\bf q})$
using the Glauber formula and the distribution
$\rho_{\rm c,h}({\bf r}_{\rm c}, {\bf r}_{\rm h})$
given by Eq. (26). Then performing integrations over the coordinates
${\bf r}_{\rm CM}$ and ${\bf r}_{\rm c}'$ (or ${\bf r}_{\rm h}'$)
we obtain:
\begin{equation}
F_{\rm c,h}({\bf q}) = {\rm exp}(- {\bf q}^2
a^2_{\rm c,h}/2A)
\cdot F_{\rm A}({\bf q}),
\end{equation}
where $F_{\rm A}({\bf q})$ is the amplitude of interest.
Consequently, the amplitude of proton-nucleus elastic scattering can
be calculated as
\begin{equation}
F_{\rm A}({\bf q}) = {\rm exp}({\bf q}^2 a^2_{\rm c,h} / 2 A) \cdot F_{\rm c,h}({\bf q}),
\end{equation}
where the amplitude $F_{\rm c,h}({\bf q})$ is given by
Eqs. (22--24).

Now let us calculate the square of the effective core size $R_{\rm c}$.
Taking into account the value of the internal core size $R^*_{\rm c}$
and equations (19) and (27) we obtain
\begin{equation}
R^2_{\rm c} = {R^*_{\rm c}}^2 + 3 \frac{a^2_{\rm c,h}}{A_{\rm c }} -
3 \frac{a^2_{\rm c,h}} {A} = {R^*_{\rm c}}^2 +
3 a^2_{\rm c,h} \frac{A_{\rm h}} {{A A_{\rm c}}}.
\end{equation}
Therefore, the parameter $a_{\rm c,h}$ used in the calculations
of the amplitude $F_{\rm c,h}({\bf q})$ is determined
through the nuclear rms radii $R_{\rm c}$ and $R^*_{\rm c}$ as
\begin{equation}
a_{\rm c,h} = [\frac{A A_c}{3 A_h} (R_{\rm c}^2 -
{R^*_{\rm c}}^2)]^{1/2}.
\end{equation}
\\
\begin{center}
3. RESULTS of CALCULATIONS\\
\end{center}

We have calculated the cross sections for proton elastic scattering
on the $^6$He and $^8$He nuclei correspondingly at the energies
717 MeV and 674 MeV in the momentum transfer range
$\sim$0 $< |t| < 0.30$ (GeV/$c$)$^2$. The values of the rms radii
$R_{\rm c}$ and $R_{\rm h}$ of the $^6$He and $^8$He nuclei and the input
parameters $\sigma_{pN}$, $\beta_{pN}$, and $\epsilon_{pN}$ of the proton-nucleon scattering amplitudes were taken from [7]:\\
$R_{\rm c} = 1.96$ fm, $R_{\rm h} = 3.30$ fm for $^6$He,\\
$R_{\rm c} = 1.81$ fm, $R_{\rm h} = 3.12$ fm for $^8$He;\\
$\sigma_{pp} = 44.6$ mb, $\beta_{pp} = 0.20$ fm$^2$, $\epsilon_{pp} = 0.069$,\\
$\sigma_{pn} =37.7$ mb, $\beta_{pn} = 0.24$ fm$^2$,
$\epsilon_{pn} = - 0.307$\\
for the proton-proton ($pp$) and proton-neutron ($pn$) interaction
in the case of $p^6$He scattering, and\\
$\sigma_{pp} = 41.9$ mb, $\beta_{pp} = 0.20$ fm$^2$, $\epsilon_{pp} = 0.129$,\\
$\sigma_{pn} = 37.4$ mb, $\beta_{pn} = 0.24$ fm$^2$,
$\epsilon_{pn} = - 0.283$\\
for the $pp$ and $pn$ interaction in the case of $p^8$He scattering.
The internal size of the core $R^*_{\rm c}$ in these nuclei
was taken as $R^*_{\rm c} =$ 1.46 fm [10]. The Coulomb interaction was
taken into account as in [11].

The results of the calculations are presented in Figs. 1 \mbox{and 2.}
The dotted, dashed and solid curves correspond respectively to the
calculations where the CM correlations were neglected, included with
approximate correction factor (7) and taken into account
exactly as was described above. Note that in these three calculations
the same nuclear one-body density distribution was used:
\begin{equation}
\rho_{\rm A} ({\bf r}) = [A_{\rm c} (3 / 2 \pi R^2_{\rm c})^{3/2}
{\rm exp} (- 3 {\bf r}^2 / 2 R^2_{\rm c}) +
A_{\rm h} (3 / 2 \pi R^2_{\rm h})^{3/2}
{\rm exp} (- 3 {\bf r}^2 / 2 R^2_{\rm h})] / A.
\end{equation}
So the difference between the results of these calculations
is due to neglect or different account of the CM correlations.
The experimental cross sections are also shown in the figures:
the hollow circles -- the data of [1], and the solid squares --
the data of [6].

As it is seen in the figures, the effect of the CM correlations
in the calculated cross sections at
$0 < |t| <0.10$ (GeV/$c$)$^2$ is rather small and in a first
approximation can be neglected. At $|t| >$ $\sim$0.10
(GeV/$c$)$^2$,
especially in the region of the first diffraction minimum and the
second diffraction maximum, the effect of the CM correlations is
rather sizeable. We also see that at high momentum transfers the
approximate approach of taking the CM correlations into account
results in a significant overestimation of the cross sections as
compared with the exact calculations.

The calculated cross sections at $|t| >$ $\sim$0.13 (GeV/$c$)$^2$
are smaller than the experimental ones. The behaviour of the
calculated cross sections at high values of $|t|$ is governed
mainly by the size of the nuclear core and its radial shape.
So, varying the core size and its shape, in principle, it is
possible to fit the calculated cross sections to the data.
In the present paper, however, we did not try to perform
such a fit. We note that the experimental cross sections of
[6] have rather large uncertainties at $|t| > 0.15$ (GeV/$c$)$^2$.
In order to get more precise sizes of the cores of the studied
nuclei and to get information on the radial shapes of the
cores, new experimental data of better quality at high $|t|$-values
and new theoretical analyses with accurate accounts of the
CM nucleon correlations are needed.\\
\begin{center}
4. CONCLUSION
\end{center}

We have shown that the effect of the CM correlations
in the cross sections for intermediate-energy proton elastic
scattering on light exotic nuclei at high momentum transfers
is rather sizeable, and it is important to take it accurately
into account. We have also shown that an approximate account
of the CM correlations with correction factor (7) is not
justified since it results in a significant overestimation
of the calculated cross sections at high $|t|$-values.\\

We are grateful to O. A. Kiselev for sending us the data of
[6] in the tabular form.

\newpage

\begin{figure}[h]
\centering\epsfig{file=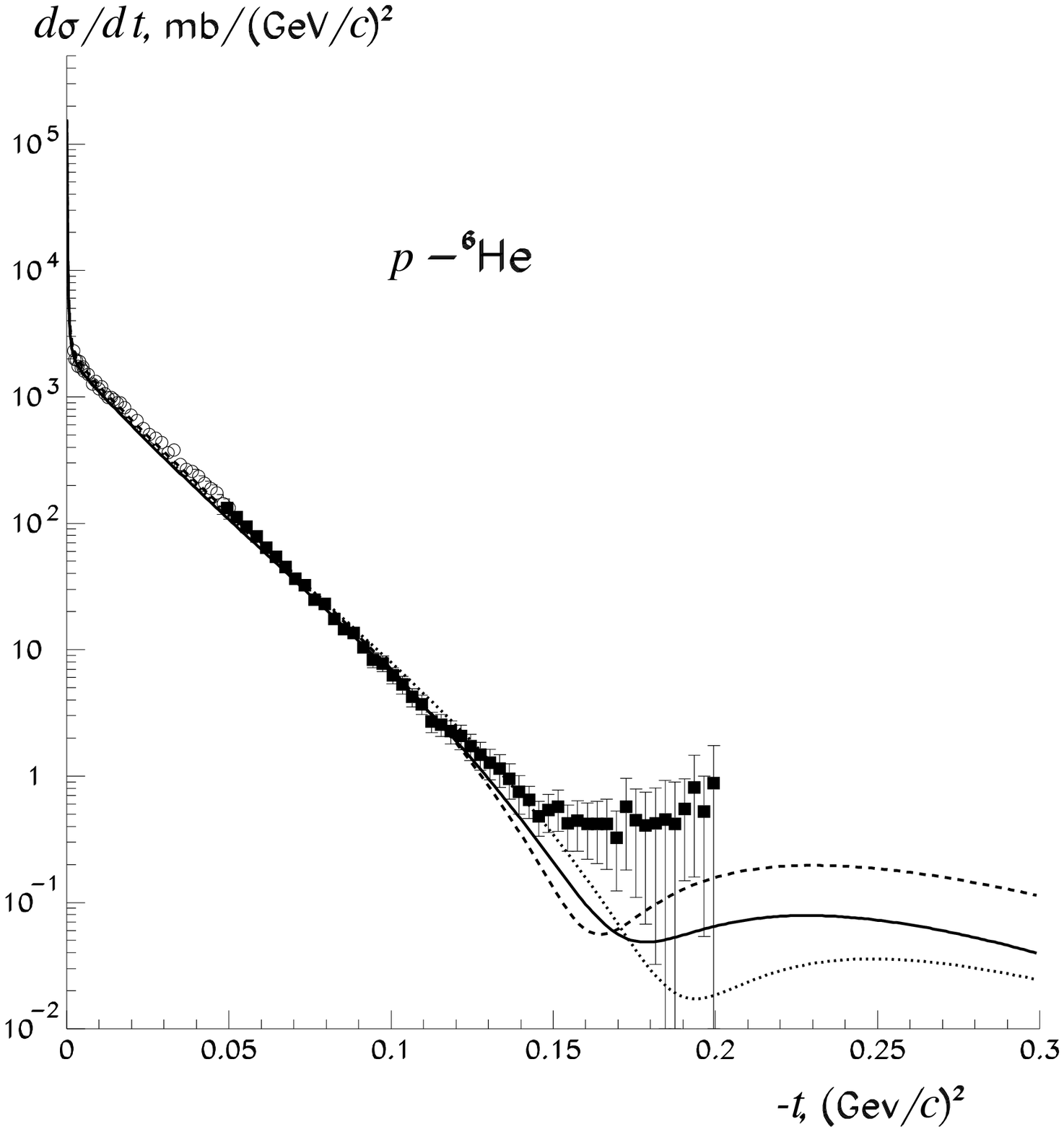,height=11cm}
\end{figure}
\noindent Fig. 1. {\small  Cross sections for proton elastic scattering
on the $^6$He nuclei at an energy of 717 MeV. Hollow circles -- data of [1], solid squares -- data of [6]. Dotted, dashed and solid curves
 correspond respectively to the cross sections calculated not taking the CM correlations into account, with the CM correlations taken into account using approximate correction factor (7), and with the CM correlations taken into account exactly}.

\newpage
\begin{figure}[h]
\centering \epsfig{file=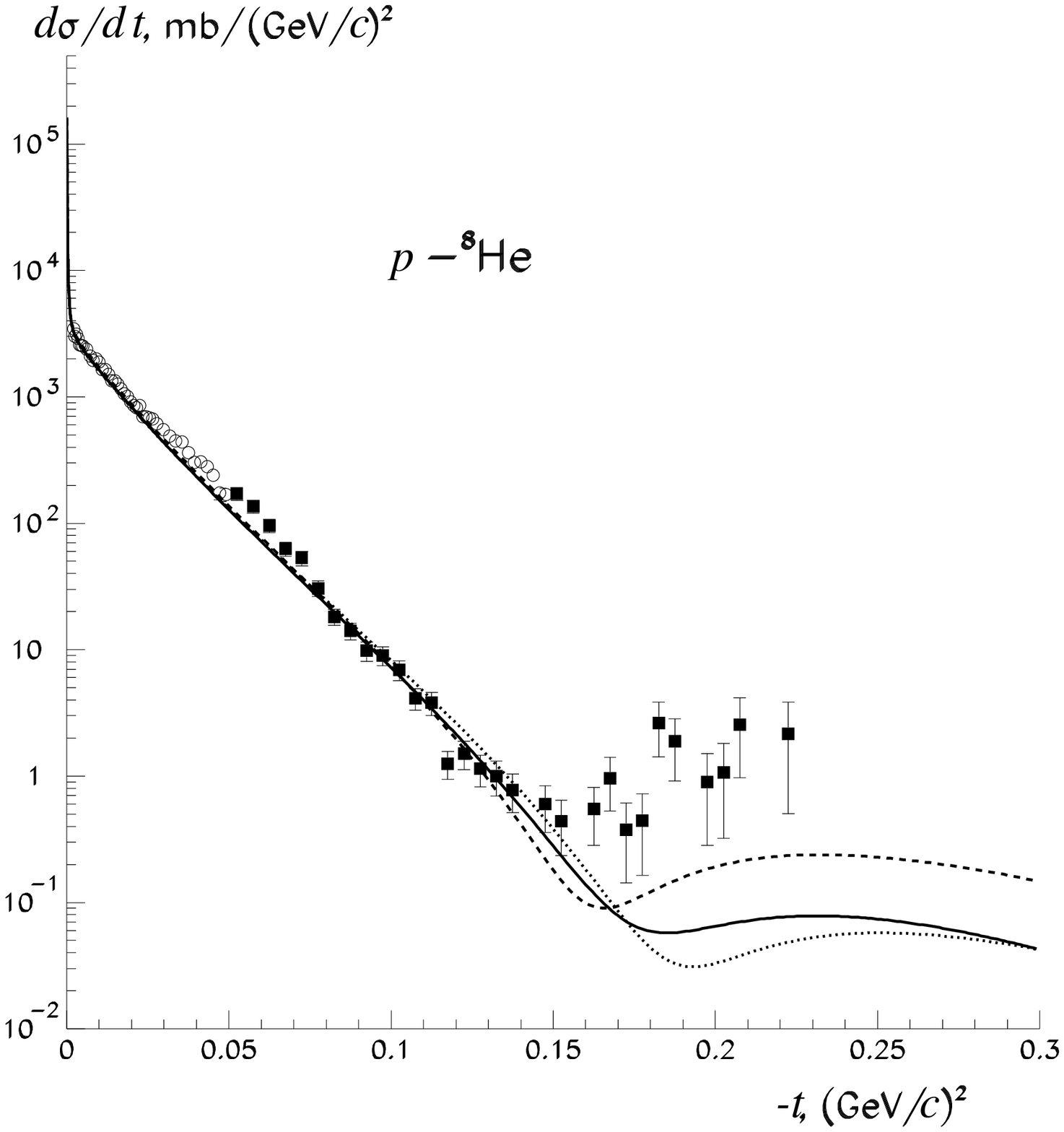,height=12.3cm}
\end{figure}
\noindent Fig. 2. {\small The same as in Fig. 1 for the
cross sections for proton elastic scattering on the $^8$He
nuclei at an energy of 674 MeV}.

\newpage
{\noindent\bf References}\\

\noindent
1. S. R. Neumaier, G. D. Alkhazov, M. N. Andronenko $et~al.$,
Nucl. Phys. A {\bf 712}, 247 (2002).\\
2. A. V. Dobrovolsky, G. D. Alkhazov, M. N. Andronenko, $et~al.$,
Nucl. Phys. A {\bf 766}, 1 (2006).\\
3. G. Ilieva, F. Aksouh, G. D. Alkhazov, $et~al.$,
Nucl. Phys. A {\bf 875}, 8 (2012).\\
4. A. A. Vorobyov, G. A. Korolev, V. A. Schegelsky, $et~al.$,
Nucl. Instr. Meth. {\bf 119}, 509 (1974).\\
5. G. D. Alkhazov, A. V. Dobrovolsky, P. Egelhof, $et~al.$,
Nucl. Phys. A {\bf 712}, 269 (2002).\\
6. O. A. Kiselev, F. Aksouh, A. Bleile $et~al.$,
Nucl. Instr. Meth. Phys. Res., Sect. A {\bf 641}, 72 (2011).\\
7. Le Xuang Chung, Oleg A. Kiselev, Dao T. Khoa, and Peter
Egelhof, Phys. Rev. C {\bf 92}, 034608 (2015).\\
8. R. J. Glauber, $Lectures~in~Theoretical~Physics$,
W.~E.~Britten and L. G. Dunham (eds.), Interscience,
New York, 1959, Vol. {\bf 1}, p. 315 .\\
9. G. D. Alkhazov, S. L. Belostotsky, A. A. Vorobyov,
Phys. Reports. {\bf 42C}, No. 2, 89 (1978).\\
10. Z.-T. Lu, P. Mueller, G. W. F. Drake, $et~al.$,
Rev. Mod. Phys. {\bf 85}, 1383 (2013). \\
11. G. D. Alkhazov, S. L. Belostotsky, O. A. Domchenkov, $et~al.$,
Nucl. Phys. A {\bf 381}, 430 (1982).
\end{document}